\documentclass[twocolumn,showpacs,preprintnumbers,amsmath,amssymb,ctexart]{revtex4}
\usepackage{pifont}
\usepackage{CJK}
\usepackage{color}
\usepackage{tipa}
\usepackage{graphicx}
\usepackage{dcolumn}
\usepackage{slashbox}
\usepackage{bm}
\usepackage{multirow}

\begin{document}
\title{Dynamical mixing between $2^3S_1$ and $1^3D_1$ charmed mesons}

\author{Hao Yu$^1$, Ze Zhao$^2$, and Ailin Zhang$^1$
\footnote{Corresponding author:zhangal@shu.edu.cn}}
\affiliation{$^1$Department of Physics, Shanghai University, Shanghai 200444, China\\
$^2$CAS Key Laboratory of Theoretical Physics, Institute of Theoretical Physics, CAS, Beijing 100190, China}
\begin{abstract}
In charmed $D$ and $D_s$ mesons sector, the matrix of a Hamiltonian in a quark potential model is computed in the $2^3S_1$ and $1^3D_1$ subspace. The masses of four mixed states of $2^3S_1$ and $1^3D_1$ denoted with $D^*_1(2635)$, $D^*_1(2739)$, $D^*_{s1}(2715)$ and $D^*_{s1}(2805)$ are obtained. It is an off-diagonal part of the spin-orbit tensor interaction that causes the mixing between the $2^3S_1$ and $1^3D_1$ states. The mixing angles between the $2^3S_1$ and $1^3D_1$ states are tiny. Under the mixing, a $^3P_0$ model is employed to compute the hadronic decay widths of all OZI-allowed decay channels of the four mixed states. The two light mixed states $D^*_1(2635)$ and $D^*_{s1}(2715)$ are close in mass to $D^*_J(2600)$ and $D^*_{s1}(2700)$, while the two heavy mixed states $D^*_1(2739)$ and $D^*_{s1}(2805)$ are lighter in mass than $D(2750)$ and $D^*_{s1}(2860)$. The mixing angles obtained from dynamical interaction are inconsistent with the mixing angles obtained from hadronic decay. Based on mass spectra and hadronic decay analyses, $D^*_J(2600)$, $D(2750)$, $D^*_{s1}(2700)$ and $D^*_{s1}(2860)$ are impossibly the mixed states of $2^3S_1$ and $1^3D_1$ at the small mixing angles. The inconsistence implies that $D^*_1(2760)$ and $D^*_{s1}(2860)$ have not been properly resolved from present experimental data, or there exist large unknown off-diagonal interactions that result in large mixing angles.
\end{abstract}

\maketitle

\section{INTRODUCTION}
$D$ and $D_s$ mesons consist of a light quark ($u$, $d$ or $s$) and a heavy $c$ quark, they behave like a hydrogen atom. These states have been studied in many models. The study of the spectrum, decay and production of $D$ and $D_s$ mesons is helpful to detect the internal quark dynamics such as the the heavy quark symmetry or the light quark chiral symmetry.

$S$-wave and $P$-wave charmed mesons ($D$ and $D_s$) without radial excitation have been well established. The higher located states are the $2S$ and $1D$ ones~\cite{pdg}, which have not been definitely identified for some reasons. In experiment, the spin and parity are difficult to determine.

$D^*(2600)$ and $D^*(2760)$ were first observed in inclusive $e^+e^-$ collisions by the BaBar Collaboration~\cite{2600 ratio} in the decay channels $D^+\pi^-$, $D^0\pi^+$ and $D^{*+}\pi^-$, where they were suggested as the $2^3S_1$ and $^3D_1$ charmed meson, respectively. In addition to their masses and widths, the branching ratios were measured
\begin{eqnarray*}
&\frac{\Gamma(D^*(2600)^0 \to D^+\pi^-)}{\Gamma(D^*(2600)^0 \to D^{*+}\pi^-)}=0.32\pm0.02\pm0.09, \\
&\frac{\Gamma(D^*(2760)^0 \to D^+\pi^-)}{\Gamma(D^*(2760)^0 \to D^{*+}\pi^-)}=0.42\pm0.05\pm0.11.
\end{eqnarray*}
The helicity angle $\Theta_H$ distributions of $D^*(2600)$ were consistent with the expectations for a natural parity ($P=(-1)^J$)~\cite{2600 ratio}.

Three years later, two resonances named $D^*_J(2650)$ and $D^*_J(2760)$ with a natural parity were observed in the $D^{*+}\pi^-$ mass spectrum in inclusive $pp$ collision by the LHCb Collaboration~\cite{2760 data1}. In this experiment, $D^*_J(2650)$ was tentatively identified as a $J^P=1^-$ radial excitation $2^3S_1$ charmed meson and $D^*_J(2760)$ was identified as a $J^P=1^-$ orbital excitation $1^3D_1$ charmed meson. Subsequently, $D^*_J(2650)$ and $D^*_J(2760)$ are believed the previously observed $D^*(2600)$ and $D^*(2760)$, respectively.

In addition to inclusive production in $e^+e^-$ and $pp$ collisions, highly excited heavy flavor resonances were also produced in exclusive $B$ decays. In exclusive B decays, $D^*_1(2760)$ was observed in the $B^-\to D^*_1(2760)^0K^-$ decay~\cite{prd91} and $D^*_3(2760)$ was observed in $B^0\to \bar D^0\pi^+\pi^-$~\cite{prd92}. The spin of $D^*_1(2760)$ was determined with $1$ through a Dalitz plot analysis~\cite{prd91}. In particular, the analysis indicates that $D^*(2760)$ observed in $e^+e^-$ and $pp$ collisions consists of $D^*_1(2760)$ and $D^*_3(2760)$~\cite{prd91,prd92} observed in B decays.

However, The observed $D^*(2760)$ in $D^+\pi^-$, $D^0\pi^+$ and $D^{*+}\pi^-$ in inclusive $e^+e^-$ and $pp$ collisions is denoted with $D(2750)$ in the charmed mesons list in PDG2018. In particular, $D(2750)$ is denoted with $D^*_3(2750)$ in a separate page. Obviously, $D^*_1(2760)$ and $D^*_3(2760)$ have not been properly resolved from present experimental data. The $D^*(2600)$ is denoted with $D^*_J(2600)$ in PDG2018. Both $D^*_J(2600)$ and $D(2750)$ are omitted from summary table in PDG2018.

$D^*_{s1}(2700)^\pm$ was first observed by BaBar~\cite{babar} and then by Belle~\cite{belle,belle1} in
$B^+\to \bar D^0D_{s1}\to\bar D^0D^0K^+$ decay with $J^P=1^-$. $D^*_{sJ}(2860)$ was first reported by BaBar~\cite{babar} in
$D_{sJ}(2860)\to D^0K^+~,~D^+K^0_s$ with a natural spin-parity. $D^*_{s1}(2700)$ and $D^*_{sJ}(2860)$ were also observed in inclusive $e^+e^-$ collision by BaBar Collaboration~\cite{Ds ratio}. Subsequently, it is found that $D^*_{sJ}(2860)$ produced in $e^+e^-$ and $pp$ collisions by BaBar and LHCb consists of $D^*_{s1}(2860)$ and $D^*_{s3}(2860)$~\cite{lhcb,lhcb1}.

Both $D^*_{s1}(2700)^\pm$ and $D^*_{s1}(2860)$ have the decay channels $DK$ and $D^*K$. The ratios of branching fractions were given in the Review of Particle Physics (2018)~\cite{pdg}
\begin{eqnarray*}
&\frac{\Gamma(D^*_{s1}(2700)^+ \to D^{*0}K^+)}{\Gamma(D^*_{s1}(2700)^+ \to D^0K^+)}=0.91\pm0.13\pm0.12, \\
&\frac{\Gamma(D^*_{s1}(2860)^+ \to D^{*0}K^+)}{\Gamma(D^*_{s1}(2860)^+ \to D^0K^+)}=1.10\pm0.15\pm0.19.
\end{eqnarray*}
The experimental results about their masses, decay widths and some branching fraction ratios are presented in Table.~\ref{tab:spectrum}

\begin{table*}[t]
\begin{center}
\caption{Experimental results of $2S$ and $1D$ candidates of D and $D_s$~\cite{pdg}.}\label{tab:spectrum}
\begin{tabular}{l  cccc  c  c  c  c }\hline \hline
~State~~~~~~ & ~~~~~~~~Experiments~~~~~~~~ &
~~~~~~~~~Mass (MeV)~~~~~~~~~ & ~~~~~Width (MeV)~~~~~ &~~~~~Branching ratios~~~~~\\
\hline
$D^*_J(2600)$ & BaBar\cite{2600 ratio},LHCb\cite{2760 data1} & $2623\pm 12$ & $139\pm 31$ & ${\Gamma(D\pi)\over \Gamma(D^*\pi)}=0.32\pm0.02\pm0.09$  \\
$D(2750)$ & BaBar~\cite{2600 ratio},LHCb~\cite{2760 data1} & $2763.5\pm 3.4$ &
$66\pm 5$ & ${\Gamma(D\pi)\over \Gamma(D^*\pi)}=0.42\pm0.05\pm0.11$ \\
\hline \hline

$D^*_{s1}(2700)$ & BaBar\cite{babar},Belle~\cite{belle,belle1} & $2708.3^{+4.0}_{-3.4}$ & $120\pm 11$ & ${\Gamma(D^*K)\over \Gamma(DK)}=0.91\pm0.13\pm0.12$ \\

$D^*_{s1}(2860)$ & BaBar\cite{babar} &  $2859\pm 12\pm 24$ & $159\pm 23\pm 77$ &
${\Gamma(D^*K)\over \Gamma(DK)}=1.10\pm0.15\pm0.19$ \\ \hline
\hline									
\end{tabular}
\end{center}
\end{table*}	

In theory, the spectroscopy of heavy-light mesons has been systematically studied in the relativized quark model~\cite{gi,gk,gm}, heavy quark symmetry theory~\cite{hqet,hqet1}, relativistic quark model~\cite{rq,rq1}, chiral quark model~\cite{pe,zhao}, lattice QCD~\cite{latt,latt1}, coupled channels models~\cite{cc1,cc2} and some other models~\cite{other1,other2,other3,other4,other5,other6,other7,other8}. More references can be found in reviews~\cite{review1,review2,review3,review4} and therein.

For low lying heavy-light mesons, theoretical predictions of the masses and the decay data are consistent with experiments. For highly excited resonances, the case is complicated. The mixing between different eigenstates may shift the predicted mass and change the decay widths. In Refs.~\cite{cornell,oka,rosner,gi}, it is noted that the mixing may arise from an internal quark dynamics or an interaction between the hadrons and their decay channels. In particular, it is pointed out that the antisymmetric piece of the spin-orbit interaction will cause a $^3L_J-^1L_J$ mixing between the mesons with unequal quark masses and the color hyperfine interaction will cause a $^3L_J-^3L'_J$ mixing~\cite{gi}.

The mixing between the $^3L_J$ and $^1L_J$ eigenstates such as the $1^1P_1-1^3P_1$ mixing has been explored in detail both through their mass spectra and through their strong decays~\cite{gi,gk,cahn,swanson}.

The mixing between the $^3L_J$ and $^3L'_J$ eigenstates such as the $2^3S_1-1^3D_1$ mixing has been explored~\cite{close,li,zhang,zhong,zhang1,chen}. In Ref.~\cite{close}, the mixing angle is determined with $\theta=-0.5$ radians from a simple masses mixing matrix of the physical states ($2.69$ GeV and $2.81$ GeV) and the predicted states of the $2^3S_1$ and $1^1D_1$ $D_s$ mesons ($2.71$ GeV and $2.78$ GeV, respectively). The mixing angle changes sign when the internal quark components of the meson are charge conjugated into their anti-quarks. Their predicted hadronic decay widths at this determined mixing angle in the $^3P_0$ model is consistent with experimental data.

In Ref.~\cite{li}, a similar mixing scheme of the $2^3S_1$ and $1^3D_1$ $D_s$ as that in Ref.~\cite{close} is employed, and the mixing angle is determined through a comparison of the predicted hadronic decay widths of the $D_s$ states in the the $^3P_0$ model with the experimental data. $1.12\leq \theta \leq 1.38$ radians (opposite in sign with opposite internal quarks) is fixed for $D^*_{s1}(2710)$, while $1.26\leq \theta \leq 1.31$ is fixed for $D_{sJ}(2860)$.

In Refs.~\cite{zhang1,chen}, the similar mixing scheme of the $2^3S_1$ and $1^3D_1$ $D$ and $D_s$ is employed. The mixing angle is studied through a comparison of the predicted hadronic decay widths of the $D_s$ states in terms of the decay formula developed by Eichten, Hill, and Quigg~\cite{hqet} with the experimental data. $\theta=4^\circ\to 17^\circ$ and $\theta=-16^\circ\to -4^\circ$ are obtained for $D^*_1(2600)$ and $D^*_{s1}(2700)$, respectively. The mixing angles are found small.

However, a dynamical exploration of the $2^3S_1-1^3D_1$ mixing has not been performed. In fact, the mixing angles determined through the mass spectra are not consistent with those determined through the decay properties. Therefore, the fixed mixing angles from experiments are different in different references. In experiment, in order to identify the $D^*_J(2600)$, $D(2750)$, $D^*_{s1}(2700)$ and $D^*_{s1}(2860)$, it is also important to systematically study the mixing between the $2^3S_1$ and $1^3D_1$ $D$ and $D_s$ mesons. For these purposes, we study the dynamical mixing between the $2^3S_1$ and $1^3D_1$ in the quark potential model firstly, and subsequently explore their strong decay in the $^3P_0$ model.

The paper is organized as follows. In the second section, the mixing mechanism between the $2^3S_1$ and $1^3D_1$ $D$ and $D_s$ mesons is explored in the quark potential model, and the mixing angles are dynamically determined. The hadronic decays of the four mixed states are explored in the $^3P_0$ model in Sec. III. In the final section, the conclusions and discussions are given.

\section{Dynamical mixing between $2^3S_1$ and $1^3D_1$}
To describe the heavy-light meson states, two kinds of eigenstates are often employed. One is the $|J,L,S\rangle$ (denoted with $^{2S}L_J$) with $J=L+S$ and $S=S_q+S_{\bar q}$ where $L$ is the orbital angular momentum, and $S_q$, $S_{\bar{q}}$ are the spins. Another one is the $|J,j\rangle$ (denoted with $j^P$), where $P$ is parity, $j=L+S_q$ is the angular momentum of light quark freedom. Physical heavy-light mesons are usually not the eigenstates $|J,L,S\rangle$ or $|J,j\rangle$, they are the mixing states of these eigenstates. Eigenstates $|J,L,S\rangle$ will be employed in the following.

In the quark potential model, the inter-quark interactions include the spin-spin interaction, the color-magnetic
interaction, the spin-orbit interaction, and the tensor force~\cite{cornell,gi,swanson}. In our analysis, the relativised quark model~\cite{swanson} is employed for our analysis, where the Hamiltonian is
\begin{gather}
H=T+V_{q\bar{q}}\\
V_{q\bar{q}}=V_{conf}+V_{SD}
\end{gather}
where $V_{conf}$ is the standard Coulomb and linear scalar interaction, the spin-orbit and color tensor interaction $V_{SD}$ is rewritten as
\begin{flalign}\label{eqvsd}
V_{SD}=&(\frac{S_q}{2m_q^2}+\frac{S_{\bar{q}}}{2m^2_{\bar{q}}})\cdot L(\frac{1}{r}\cdot \frac{dV_{conf}}{dr}+\frac{2}{r}\cdot \frac{dV_1}{dr})\nonumber\\
+&\frac{(S_q+S_{\bar q})\cdot L}{m_qm_{\bar{q}}}(\frac{1}{r}\cdot \frac{dV_2}{r})+\frac{3S_q\cdot \hat{r}S_{\bar{q}}\cdot \hat{r}-S_q\cdot S_{\bar{q}}}{3m_qm_{\bar{q}}}\cdot V_3\nonumber\\
+&[(\frac{S_q}{m_q^2}-\frac{S_{\bar{q}}}{m_{\bar{q}}^2})+\frac{S_q-S_{\bar{q}}}{m_qm_{\bar{q}}}]\cdot L V_4\nonumber\\
+&\frac{32\alpha_s\sigma^3e^{-\sigma^2r^2}}{9\sqrt{\pi}m_qm_{\bar{q}}}S_q\cdot S_{\bar{q}}
\end{flalign}
The explicit form of $V_1$, $V_2$, $V_3$ and $V_4$ are~\cite{swanson,ef}
\begin{flalign}
	V_1(m_q,m_{\bar{q}},r)=&-br-C_F \frac{1}{2r} \frac{\alpha_s^2}{\pi}(C_F \nonumber\\
	&-C_A(ln[(m_qm_{\bar{q}})^{1/2}r]+\gamma_E)) \nonumber\\
	V_2(m_q,m_{\bar{q}},r)=&-\frac{1}{r}C_F\alpha_s[1+\frac{\alpha_s}{\pi}[\frac{b_0}{2}[ln(\mu r)+\gamma_E] \nonumber\\
	&+\frac{5}{12}b_0-\frac{2}{3}C_A+\frac{1}{2}(C_F \nonumber\\
	&-C_A(ln[(m_qm_{\bar{q}})^{1/2}r]+\gamma_E))]] \nonumber\\
	V_3(m_q,m_{\bar{q}},r)=&\frac{3}{r^3}C_F\alpha_s[1+\frac{\alpha_s}{\pi}[\frac{b_0}{2}[ln(\mu r)+\gamma_E-\frac{4}{3}]\nonumber\\
	&+\frac{5}{12}b_0-\frac{2}{3}C_A+\frac{1}{2}(C_A+2C_F \nonumber\\
	&-2C_A(ln[(m_qm_{\bar{q}})^{1/2}r]+\gamma_E-\frac{4}{3}))]] \nonumber\\
	V_4(m_q,m_{\bar{q}},r)=&\frac{1}{4r^3}C_FC_A\frac{\alpha_s^2}{\pi}ln\frac{m_{\bar{q}}}{m_q}
\end{flalign}
with $C_F=\frac{4}{3}$, $C_A=3$, $b_0=9$, and $\gamma_E=0.5772$. The model parameters are $\alpha_s =0.53$, $\mu =1.0$, $\sigma =1.13$, $b=0.135$, $C_{c\bar{u}}=-0.305$, and $C_{c\bar{s}}=-0.254$, they were given in Ref.~\cite{swanson}. The quak masses are chosen as following: $m_c=1450$ MeV, $m_u=m_d=450$ MeV, and $m_s=550$ Me V. In term of these parameters, the predicted masses of the $1S$ and $1P$ $D$ and $D_s$ mesons agree well to the experimental data, which are presented in Table.~\ref{D mesons} and Table.~\ref{Ds mesons}

\begin{table}[]
	\caption{Masses of $1S$ and $1P$ $D$ meson (MeV)}\label{D mesons}
	\begin{tabular}{p{0cm}p{3.0cm}p{3.0cm}p{2.0cm}}
		\hline
		& State                            & This Work      & PDG                       \\
		\hline
		&$1^1S_0$                          &  $1867 $   & $1869  $                      \\
		&$1^3S_1$                          &  $2017 $   & $2010  $                      \\
		&$1^3P_0$                          &  $2257 $   & $2308  $                      \\
		&$1^3P_2$                          &  $2473 $   & $2460  $                      \\
		&$1P$                              &  $2399 $   & $2422  $                      \\
		&$1P^{\prime}$                     &  $2429 $   & $2427  $                      \\
		\hline
	\end{tabular}
\end{table}

\begin{table}[]
	\caption{Masses of $1S$ and $1P$ $D_s$ meson (MeV)}\label{Ds mesons}
	\begin{tabular}{p{0cm}p{3.0cm}p{3.0cm}p{2.0cm}}
		\hline
		& State                            & This Work      & PDG                       \\
		\hline
		&$1^1S_0$                          &  $1969 $   & $1969  $                      \\
		&$1^3S_1$                          &  $2114 $   & $2112  $                      \\
		&$1^3P_0$                          &  $2353 $   & $2317  $                      \\
		&$1^3P_2$                          &  $2567 $   & $2572  $                      \\
		&$1P$                              &  $2494 $   & $2459  $                      \\
		&$1P^{\prime}$                     &  $2517 $   & $2535  $                      \\
		\hline
	\end{tabular}
\end{table}

As well known, the $H$ is not diagonal in the basis $|J,L,S\rangle$ or $|J,j\rangle$. The relation between $|J,L,S\rangle$ and $|J,j\rangle$ can be found in Refs.~\cite{gm,cahn}. From Ref.~\cite{gm}, the off-diagonal interaction arises from the tensor interaction
\begin{flalign}
	V_{tensor}=\frac{3S_q\cdot \hat{r}S_{\bar{q}}\cdot \hat{r}-S_q\cdot S_{\bar{q}}}{3m_qm_{\bar{q}}}\cdot V_3(r)
\end{flalign}
which can be written in an irreducible representation as
\begin{flalign}
	V_{tensor}=6\sqrt{\frac{8\pi}{15}}Y^{(2)}\cdot S^{(2)}\cdot V_3(r)\nonumber
\end{flalign}
where $Y^{(2)}$ is a rank $2$ spherical harmonics and $S^{(2)}=(S_q^{(1)} \times S_{\bar{q}}^{(1)})^{(2)}$ with spin operator $S_q^{(1)}$, $S_{\bar{q}}^{(1)}$ in the spherical basis.

The matrix element of the tensor term is obtained through the Wigner-Eckhart theorem~\cite{WE},
\begin{flalign}
	&\left\langle J,L,S|V_{tensor}|J,L^{\prime},S \right\rangle\nonumber\\
	&=(-1)^{L+S+J}
	\begin{Bmatrix}
		S & 2 & S\\
		L & J & L^{\prime}\\		
	\end{Bmatrix}
\left\langle L||Y^{(2)}||L^{\prime} \right\rangle\ \left\langle S||S^{(2)}||S \right\rangle \nonumber\\
&\times\left\langle J,L,S|V_3(r)|J,L^{\prime},S \right\rangle\ \nonumber
\end{flalign}
where $\left\langle L||Y^{(2)}||L^{\prime} \right\rangle\ $ is a space reduced matrix element
\begin{flalign}
    &\left\langle L||Y^{(2)}||L^{\prime} \right\rangle\ \nonumber\\
    &= (-1)^L\sqrt{\frac{5(2L+1)(2L^{\prime}+1)}{4\pi}}\times
    \begin{pmatrix}
    	L & 2 & L^{\prime}\\
    	0 & 0 & 0\\		
    \end{pmatrix}\nonumber
\end{flalign}
and $\left\langle S||S^{(2)}||S \right\rangle $ is the spin reduced matrix element which is $\frac{\sqrt{5}}{2}$ at $S=1$.

In the subspace of $\langle 2^3S_1|$ and $\langle 1^3D_1|$, the non-diagonal matrix of the Hamiltonian is
\begin{flalign}
	\begin{bmatrix}
		{H_{11}} & {H_{12}}\nonumber\\
		\\
		{H_{21}} & {H_{22}}\\
	\end{bmatrix}.
\end{flalign}
The numerical matrix of H in the subspace of $\langle 2^3S_1|$ and $\langle 1^3D_1|$ for $D$ and $D_s$ mesons are
\begin{flalign}
\left[
\begin{array}{c}
2635.16 \quad -0.21\\
\\
-0.21 \quad 2738.51
\end{array}
\right]\rm {and}\left[\begin{array}{c}
2714.76  \quad -0.29\\
\\
-0.29 \quad 2805.49   \\
\end{array}
\right]
\end{flalign}
, respectively.

Without the off-diagonal tensor interaction, $\langle 2^3S_1|$ and $\langle 1^3D_1|$ are the eigenstates of the left $H$. In this case, the eigenvalues of the $\langle 2^3S_1|$ and $\langle 1^3D_1|$ $D$ mesons are $2635.16$ MeV and $2738.51$ MeV, respectively. The eigenvalues of the $\langle 2^3S_1|$ and $\langle 1^3D_1|$ $D_s$ mesons are $2714.76$ MeV and $2805.49$ MeV, respectively. The masses of $\langle 2^3S_1|$ charmed mesons are comparable to those in Ref.~\cite{gm}, but the masses of $\langle 1^3D_1|$ charmed states are lower than those in the same reference.

When the light and heavy mixed sates are denoted with $|D^{*L}_1\rangle$ and $|D^{*H}_1\rangle$~\cite{li,chen}, respectively, the matrix H can be diagolized in the physical states (mixed states)
\begin{flalign*}
\left[
\begin{array}{c}
|D^{*L}_1\rangle\\
\\
|D^{*H}_1\rangle
\end{array}
\right]=\left[\begin{array}{c}
cos\theta  \quad  sin\theta\\
\\
-sin\theta  \quad cos\theta   \\
\end{array}
\right]\left[
\begin{array}{c}
|2^3S_1\rangle\\
\\
|1^3D_1\rangle
\end{array}
\right]
\end{flalign*}
with a mixing angle $\theta$. After diagolization, $H$ is turned into~\cite{cahn}
\begin{flalign}
&\left[\begin{array}{c}
H^\prime_{11} \quad {0}\\
\\
{0} \quad {H^\prime_{22}}\\
\end{array}
\right]=\nonumber\\
&\left[
\begin{array}{c}
cos\theta  \quad  sin\theta\\
\\
-sin\theta  \quad cos\theta\\
\end{array}
\right]\left[\begin{array}{c}
H_{11} \quad {H_{12}}\\
\\
{H_{21}} \quad {H_{22}}\\
\end{array}
\right]\left[
\begin{array}{c}
cos\theta  \quad  sin\theta\\
\\
-sin\theta  \quad cos\theta\\
\end{array}
\right]^{-1}
\end{flalign}
where $H'_{11}$ and $H'_{22}$ are the energy eigenvalues of the physical $D^{*L}_1$ and $D^{*H}_1$ states, respectively.

With previous formulas in hand, we obtain the masses of the light and heavy mixed physical states and the mixing angles as follows

\begin{eqnarray*}
	&M(D^{*L}_1)= 2635.16~\rm {MeV},\\
	&M(D^{*H}_1)= 2738.51~\rm {MeV},\\
	&\theta_{c\bar{q}}\approx 0.12^{\circ}.\\
	&M(D^{*L}_{s1})= 2714.76~\rm {MeV},\\
	&M(D^{*H}_{s1})= 2805.49~ \rm {MeV},\\
	&\theta_{c\bar{s}}\approx 0.18^{\circ}.\\
\end{eqnarray*}
These four mixed states will be denoted with $D^*_1(2635)$, $D^*_1(2739)$, $D^*_{s1}(2715)$ and $D^*_{s1}(2805)$ throughout this paper. Obviously, the mixing angles between the $2^3S_1$ and $1^3D_1$ for $D$ and $D_s$ are very small, and the off-
diagonal interactions resulting from the tensor interaction almost do not change the eigenvalues.

From Table.~\ref{tab:spectrum}, the masses of the two light mixed $D^*_1(2635)$ and $D^*_{s1}(2715)$ are close to the masses of $D^*_J(2600)$ and $D^*_{s1}(2700)$, but the masses of the two heavy mixed $D^*_1(2739)$ and $D^*_{s1}(2805)$ are lighter than the masses of $D(2750)$ and $D^*_{s1}(2860)$.

Obviously, an off-diagonal tensor interactions inversely proportional to the products of heavy quark and light quark mass in Eq.~(\ref{eqvsd}) results in a tiny mixing, and the heavy mixed $D^{*H}_1$ states have masses lighter than $D(2750)$ and $D^*_{s1}(2860)$.

There are two possibilities that may result in lighter masses of $D^{*H}_1$ in comparison to $D(2750)$ and $D^*_{s1}(2860)$. First, $D(2750)$ and $D^*_{s1}(2860)$ have not been definitely identified. As analyzed in Refs.~\cite{prd91,prd92,lhcb,lhcb1,gm,zhao1}, $D^*(2760)$ ($D(2750)$) observed in $e^+e^-$ and $pp$ collisions was resolved into the two $D^*_1(2760)$ and $D^*_3(2760)$ $D$ states, $D^*_{sJ}(2860)$ observed in $e^+e^-$ and $pp$ collisions was also resolved into the two $D^*_{s1}(2860)$ and $D^*_{s3}(2860)$ $D_s$ states. However, $D(2750)$ and $D^*_{s1}(2860)$ were observed in inclusive $e^+e^-$ and $pp$ collisions with a natural parity, but the spin and parity are difficult to determine in those inclusive decays. $D^*_{s1}(2860)$ and $D^*_{s3}(2860)$ $D_s$ were observed and measured with definite spin in the exclusive B decays~\cite{prd91,prd92}. Obviously, the analyses of the resolve are not sufficient.
In PDG2018, $D(2750)$ was simply denoted with $D^*_3(2750)$ and $D^*_1(2760)$ is missing. In other words, the fixed data of $D^*(2760)$ and $D^*_{sJ}(2860)$ are not sufficient to give the right data of $D^{*H}_1$ and $D^{*H}_{s1}$. In experiment, it is important to figure out proper ways to give the exact masses and decay widths of the resolved $D^{*H}_1$ and $D^{*H}_{s1}$ through $D^*(2760)$ ($D(2750)$) and $D^*_{sJ}(2860)$ in the future.

Secondly, if there exists any other unknown interaction in the Hamiltonian which may result in a large mixing between the $2^3S_1$ and $1^3D_1$ for $D$ and $D_s$, the theoretical predictions of the masses will be consistent with experiments. In order to see how the masses of the four mixed states depend on the mixing angles, the variation of their masses with the mixing angles is plotted in Fig. 1. In a large range of the mixing angles, the masses of $D^{*H}_1$ and $D^{*H}_{s1}$ turn larger with larger mixing angles, while the masses of $D^{*L}_1$ and $D^{*L}_{s1}$ turn smaller with larger mixing angles.

\begin{figure}[ht]
\centering
\includegraphics[scale=0.3]{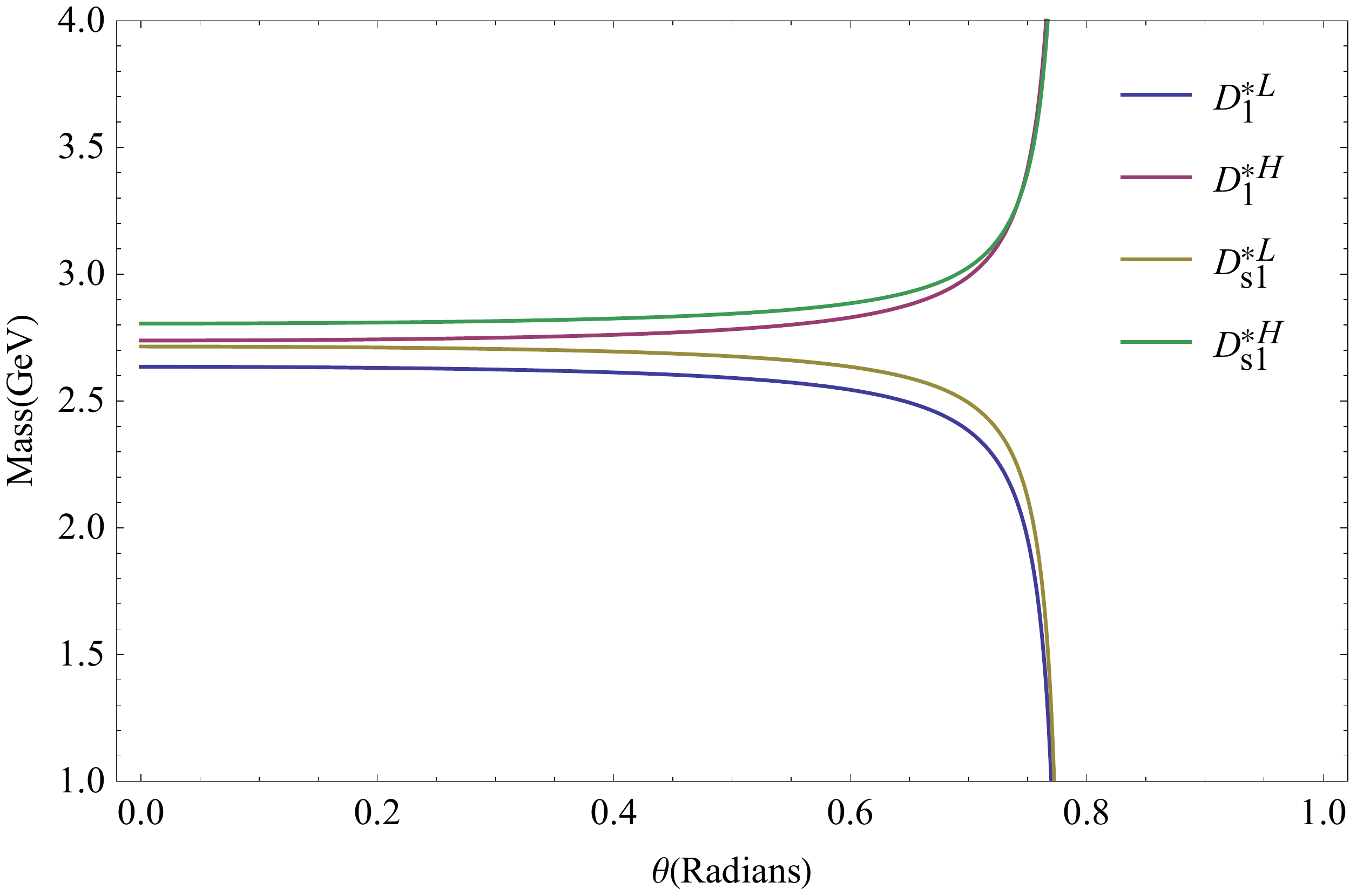}
\caption{Masses of the four mixed mesons with the mixing angles.}
\end{figure}

In Ref.~\cite{close}, the $D^*_{sJ}(2860)$ was regarded as the $D^{*H}_{s1}$, and a large mixing angle $\theta=-0.5$ radians has been phenomenologically obtained, but how the large mixing results from has not been studied. Whether there is an unknown interaction that can result in a large mixing between the $2^3S_1$ and $1^3D_1$ charm mesons requires more exploration. For this purpose, an accurate measurement of the masses of $D^{*L}_1$ ($D^{*L}_{s1}$) and $D^{*H}_1$ ($D^{*H}_{s1}$) in the meantime is very important.

\section{Hadronic decay of $D^*_1(2635)$, $D^*_1(2739)$, $D^*_{s1}(2715)$ and $D^*_{s1}(2805)$}
In order to learn the internal quark dynamics, another way is to study the strong decay of hadrons. In the case of $2^3S_1$ and $1^3D_1$ mixing, the hadronic decay of the four mixed states are explored in the $^3P_0$ model in this section.

As well known, the $^3P_0$ model is usually called as the quark-pair creation model. It has been employed extensively to study the Okubo-Zweig-Iizuka(OZI)-allowed hadronic decay processes. The model was first proposed by Micu~\cite{micu1969} and developed by Yaouanc et al~\cite{yaouanc1,yaouanc2,yaouanc3}. The "QCD" decay mechanism of the $^3P_0$ model was studied in references~\cite{swanson1,swanson2,swanson3}. The $^3P_0$ model is fundamentally based on a flux tube picture of the quark confinement. Based on the flux tube picture of the quark confinement, the strong decay and $p\bar p$ annihilation processes are also well described in a $^3S_1$ model~\cite{zpc84,npa84,plb93,prc04}.

In the $^3P_0$ model, the decay of a meson takes place through a $q\bar{q}$ pair creation with the vacuum quantum number $J^{PC}=0^{++}$. The hadronic partial decay width $\Gamma$ of a decay process $A\rightarrow B+C$
\begin{eqnarray}
\Gamma  = \pi ^2 \frac{|\vec{k}|}{m_A^2}\sum_{JL} |{\mathcal{M}^{JL}}|^2
\end{eqnarray}
where $|\vec{k}|=\frac{{\sqrt {[m_A^2-(m_B-m_C )^2][m_A^2-(m_B+m_C)^2]}}}{{2m_A }}$ is the momentum of the final states B and C in the initial meson A's center-of-mass frame, and $\mathcal{M}^{JL}$ is the partial wave amplitude of $A \rightarrow B+C$.

For mixed states $|D^{*L}_1\rangle$ and $|D^{*H}_1\rangle$ with mixing angle $\theta$,
\begin{flalign}
	&\Gamma(\left|D_L \right\rangle) \nonumber\\
	&=\pi^2\frac{\left|\vec{K} \right|^2 }{m_A^2}\sum\limits_{JL}\left|\cos\theta \mathcal{M}^{JL}(2^3S_1)-\sin\theta \mathcal{M}^{JL}(1^3D_1) \right|^2\nonumber\\
	&\Gamma(\left|D_H \right\rangle)\nonumber\\
	&=\pi^2\frac{\left|\vec{K} \right|^2 }{m_A^2}\sum\limits_{JL}\left|\sin\theta \mathcal{M}^{JL}(2^3S_1)+\cos\theta \mathcal{M}^{JL}(1^3D_1) \right|^2.
\end{flalign}

In terms of the Jacob-Wick formula, $\mathcal{M}^{JL}$ can be written as~\cite{JW},
\begin{flalign}
\mathcal{M}^{JL} (A \to BC) &= \frac{{\sqrt {2L + 1} }}{{2J_A  + 1}} \nonumber \\
&\times\sum_{M_{J_B } ,M_{J_C } } \langle {L0JM_{J_A } } |{J_A M_{J_A } }\rangle  \nonumber \\
&\times\langle {J_B M_{J_B } J_C M_{J_C } } |J, {JM_{J_A } } \rangle \nonumber \\
 &\times \mathcal{M}^{M_{J_A } M_{J_B } M_{J_C } } (\vec{K})
\end{flalign}
where $\vec{J}=\vec{J_B}+\vec{J_C}$, $\vec{J_A}=\vec{J_B}+\vec{J_C}+\vec{L}$ and $M_{J_A}=M_{J_B}+M_{J_C}$. The $\mathcal{M}^{M_{J_A } M_{J_B } M_{J_C }}$ is the helicity amplitude
\begin{flalign}
 &\mathcal{M}^{M_{J_A } M_{J_B } M_{J_C }}\nonumber \\
 &=\sqrt {8E_A E_B E_C } \gamma \sum_{\mbox{\tiny$\begin{array}{c}
M_{L_A } ,M_{S_A } ,\\
M_{L_B } ,M_{S_B } ,\\
M_{L_C } ,M_{S_C } ,m\end{array}$}}  \langle {L_A M_{L_A } S_A M_{S_A } }| {J_A M_{J_A } }\rangle \nonumber \\
 &\times\langle L_B M_{L_B } S_B M_{S_B }|J_B M_{J_B } \rangle \langle L_C M_{L_C } S_C M_{S_C }|J_C M_{J_C }\rangle\nonumber \\
 & \times \langle {1m;1 - m}|{00} \rangle\langle \chi _{S_B M_{S_B }}^{13} \chi _{S_C M_{S_C } }^{24}|\chi _{S_A M_{S_A } }^{12} \chi _{1 - m}^{34}\rangle \nonumber \\
&\times\langle\varphi _B^{13} \varphi _C^{24}|\varphi _A^{12}\varphi _0^{34} \rangle I_{M_{L_B } ,M_{L_C } }^{M_{L_A },m} (\vec{K})
\end{flalign}
where $\gamma$ is the pair-production strength constant. The detail of the flavor matrix element $\langle\varphi _B^{13} \varphi _C^{24}|\varphi _A^{12}\varphi _0^{34} \rangle$, the spin matrix element $\left\langle {\chi _{S_B M_{S_B } }^{13} \chi _{S_C M_{S_C } }^{24} } |{\chi _{S_A M_{S_A } }^{12} \chi _{1 - m}^{34} } \right\rangle$ and the momentum integral $I_{M_{L_B },M_{L_C}}^{M_{L_A },m}(\vec{K})$ can be found in Ref.~\cite{zhao1}.

In the $^3P_0$ model, numerical results depend on the parameters such as $\gamma$, the harmonic oscillator parameter $\beta$ and the constituent quark masses. $\gamma=6.947$ ($\sqrt{96\pi}$ times as the $\gamma=0.4$ in Ref.~\cite{gm}) in Refs.~\cite{zhao1,3040,gamma6.94} is also employed in this paper. For strange quark-pair $s\bar{s}$ creation, $\gamma_{s\bar{s}}=\gamma/\sqrt{3}$~\cite{yaouanc2}. The $\beta$ are taken from Ref.~\cite{diagonal matrix element}. The constituent quark masses are chosen as $m_c=1450$ MeV, $m_u=m_d=450$ MeV, and $m_s=550$ MeV~\cite{diagonal matrix element}.

In our computation, the masses of related mesons are input as follows: $m_{\pi^0}=134.977$ MeV, $m_{\pi^\pm}=139.570$ MeV, $m_{K^0}=497.611$ MeV, $m_{K^\pm}=493.677$ MeV, $m_{\rho(770)^0}=775.26$ MeV, $m_{\rho(770)^\pm}=775.11$ MeV, $m_{\eta}=547.862$ MeV, $m_{\omega}=782.65$ MeV, $m_{K^*(892)^0}=895.81$ MeV, $m_{K^*(892)^\pm}=891.66$ MeV, $m_{D^0}=1864.84$ MeV, $m_{D^\pm}=1869.61$ MeV, $m_{D^{*0}}=2006.97$ MeV, $m_{D^{*\pm}}=2010.27$ MeV, $m_{D(2550)^{0}}=2539.4$ MeV, $m_{D_1(2420)^0}=2421.4$ MeV, $m_{D_1(2420)^\pm}=2423.2$ MeV, $m_{D_1(2430)^{0,\pm}}=2427.0$ MeV, $m_{D_2^*(2460)^0}=2462.6$ MeV, $m_{D_2^*(2460)^\pm}=2464.3$ MeV, $m_{D_s^\pm}=1968.3$ MeV, $m_{D_s^{*\pm}}=1968.3$ MeV. The masses of the four mixed states are chosen as: $m_{D^*_1(2635)^0}=2635.16$ MeV, $m_{D^*_1(2739)^0}=2738.51$ MeV, $m_{D^*_{s1}(2715)}=2714.76$ MeV, $m_{D^*_{s1}(2805)}=2805.49$ MeV~\cite{pdg}.

\subsection{$D^*_1(2635)$ and $D^*_1(2739)$}

$D^*_1(2635)$ and $D^*_1(2739)$ are mixed states of $2^3S_1$ and $1^3D_1$ $D$ mesons with mixing angle $\theta=0.12^\circ$, possible hadronic decay channels and relevant partial decay widths are presented in Table.~\ref{decay1}.
\begin{table}[]
\caption{Hadronic decay widths of $D^*_1(2635)^0$ and $D^*_1(2739)^0$ as mixed states of $2^3S_1$ and $1^3D_1$ with mixing angle $\theta=0.12^\circ$ (in MeV).}\label{decay1}
\begin{tabular}{p{0cm}p{3.0cm}p{3.0cm}p{2.0cm}}
   \hline\hline
  &                                      &$D^*_1(2635)$ &$D^*_1(2739)$\\
   \hline

   & Channels                            & Width      & Width                       \\
   \hline
   &$D_1(2420)^0\pi^0$                   &  $1.46 $   & $42.88  $                      \\
   &$D_1(2420)^+\pi^-$                   &  $2.79 $   & $85.51  $                      \\
   &$D_1(2430)^0\pi^0$                   &  $6.91 $   & $7.73   $                      \\
   &$D_1(2430)^+\pi^-$                   &  $13.62 $  & $15.78   $                      \\
   &$D^0\pi^0$                           &  $0.09 $   & $18.06  $                      \\
   &$D^+\pi^-$                           &  $0.13 $   & $36.52  $                      \\
   &$D_s^+K^-$                           &  $0.25 $   & $12.51  $                      \\
   &$D^0\eta^0 $                         &  $0.34 $   & $12.11  $                      \\
   &$D_2^*(2460)^0\pi^0$                 &  $0.01 $   & $0.32   $                      \\
   &$D_2^*(2460)^+\pi^-$                 &  $0.02 $   & $0.58   $                      \\
   &$D^{*0}\pi^0$                        &  $2.36 $   & $9.95   $                      \\
   &$D^{*+}\pi^-$                        &  $4.90$    & $20.02  $                      \\
   &$D^{*0}\eta^0$                       &  $1.62 $   & $5.01   $                      \\
   &$D_s^{*+}K^-$                        &  $0.34 $   & $3.74   $                      \\
   &$D(2550)^0\pi^0$                     &  $\times$   & $0.02   $                      \\
   &$D(2550)^+\pi^-$                     &  $\times$   & $0.03   $                      \\
   &$D^0\rho^0$                          &  $\times$   & $7.29   $                      \\
   &$D^+\rho^-$                          &  $\times$   & $13.91   $                      \\
   &$D^{*0}\omega^0$                     &  $\times$   & $6.80   $ \\
   &$\Gamma_{total}$                     &  $34.84$    & $298.77   $     \\
         \hline\hline
\end{tabular}
\end{table}
From this table, the total hadronic decay widths of $D^*_1(2635)$ and $D^*_1(2739)$ are $34.84$ MeV and $298.77$ MeV, respectively. These total decay widths are largely different with the observed states'.

The following ratios are also obtained

\begin{eqnarray*}
&\frac{\Gamma(D^*_1(2635)^0 \to D^+\pi^-)}{\Gamma(D^*_1(2635)^0 \to D^{*+}\pi^-)}=0.03 \\
&\\
&\frac{\Gamma(D^*_1(2635)^0 \to D^+_sK^-)}{\Gamma(D^*_1(2635)^0 \to D^{*+}_sK^-)}=0.74  \\
&\\
&\frac{\Gamma(D^*_1(2739)^0 \to D^+\pi^-)}{\Gamma(D^*_1(2739)^0 \to D^{*+}\pi^-)}=1.82 \\
&\\
&\frac{\Gamma(D^*_1(2739)^0 \to D^+_sK^-)}{\Gamma(D^*_1(2739)^0 \to D^{*+}_sK^-)}=3.34
\end{eqnarray*}

Obviously, the obtained branching ratios $\Gamma(D^+\pi)/\Gamma(D^{*+}\pi^-)$ of $D^*_1(2635)$ is smaller than the observed one of $D^*_J(2600)$, while the branching ratios $\Gamma(D^+\pi)/\Gamma(D^{*+}\pi^-)$ of $D^*_1(2739)$ are larger than the observed one of $D(2750)$. Therefore, even if $D(2750)$ is a $J^P=1^-$ (instead of $J^P=3^-$) charmed meson, $D^*_J(2600)$ and $D(2750)$ are impossible to be identified with the combination of $2^3S_1$ and $1^3D_1$ $D$ mesons at a small mixing angle $\theta=0.12^\circ$. That is to say, the mixing angle obtained from internal quark dynamics is inconsistent with the mixing angle obtained from strong decay even if the observed $D^*_J(2600)$ and $D(2750)$ have been identified as the $D^{*L}_1$ and $D^{*H}_1$.

\subsection{$D^*_{s1}(2715)$ and $D^*_{s1}(2805)$}

$D^*_{s1}(2715)$ and $D^*_{s1}(2805)$ are mixed states of $2^3S_1$ and $1^3D_1$ $D_s$ with mixing angle $\theta=0.18^\circ$, possible hadronic decay channels and relevant partial decay widths are presented in Table.~\ref{decay2}.

\begin{table*}[]
	\caption{Hadronic decay widths of $D^*_{s1}(2715)^+$ and $D^*_{s1}(2805)^+$ as mixed states of $2^3S_1$ and $1^3D_1$ with mixing angle $\theta=0.18^\circ$ (in MeV)}\label{decay2}
	\begin{tabular}{p{0cm}p{3.0cm}p{3.0cm}p{2.0cm}}
		\hline\hline
		&                                      &$D^*_{s1}(2713)^+$ &$D^*_{s1}(2773)^+$\\
		\hline
		
		& Channels                            & Width      & Width                       \\
		\hline
		&$D^+K^0$                             &  $1.79  $   & $51.79  $                      \\
		&$D^0K^+$                             &  $1.63  $   & $51.30  $                      \\
		&$D^{*+}K^0$                          &  $17.23  $   & $26.43  $                      \\
		&$D^{*0}K^+$                          &  $17.18  $   & $26.47  $                      \\
		&$D_S^+\eta ^0$                       &  $0.50  $   & $10.40  $                      \\
		&$D_S^{*+}\eta^0$                     &  $0.94  $   & $3.36  $                      \\
		&$D^0K^{*+}$                          &  $\times $    & $8.36 $                      \\
		&$D^+K^{*0}$                          &  $\times $    & $6.52 $                      \\
		&$\Gamma_{total}$                          &  $39.27 $    & $184.63 $                      \\	
		\hline\hline
	\end{tabular}
\end{table*}
From this table, the total hadronic decay width ($39.27$ MeV) of $D^*_{s1}(2715)$ is much smaller than the observed one of $D^*_{s1}(2700)$, while the total hadronic decay width ($184.63$ MeV) of  $D^*_{s1}(2805)$ is comparable to that of $D^*_{s1}(2860)$.

The obtained ratios

\begin{eqnarray*}
&\frac{\Gamma(D^*_{s1}(2715)^+ \to D^0K^+)}{\Gamma(D^*_{s1}(2715)^+ \to D^{*0}K^+)}=0.09 \\
&\\
&\frac{\Gamma(D^*_{s1}(2805)^+ \to D^0K^+)}{\Gamma(D^*_{s1}(2805)^+ \to D^{*0}K^+)}=1.94
\end{eqnarray*}
are largely different with the observed ones of $D^*_{s1}(2700)$ and $D^*_{s1}(2860)$.

Obviously, $D^*_{s1}(2700)$ and $D^*_{s1}(2860)$ are impossible to be identified with the combination of $2^3S_1$ and $1^3D_1$ $D_s$ mesons at a mixing angle $\theta=0.18^\circ$ either. In other words, the mixing angle obtained from internal quark dynamics is inconsistent with the mixing angle obtained from strong decay either if $D^*_{s1}(2700)$ and $D^*_{s1}(2860)$ have been identified in their present data.

\section{CONCLUSIONS AND DISCUSSIONS}
In this paper, the masses of $1S$, $1P$, $1D$ and $2S$ states of $D$ and $D_s$ have been calculated in the quark potential model. The off-diagonal tensor interactions resulting in the mixing between $2^3S_1$ and $1^3D_1$ charmed mesons are computed. The mixing angles are found tiny, and the mass difference between the light $q$ quark and the $s$ quark changes the mixing angle little. Four mixed $D^{*L}_1$, $D^{*H}_1$, $D^{*L}_{s1}$ and $D^{*H}_{s1}$ are found: $D^*_1(2635)$, $D^*_1(2739)$, $D^*_{s1}(2715)$ and $D^*_{s1}(2805)$, whose masses are $2635$ MeV, $2739$ MeV, $2715$ MeV and $2805$ MeV, respectively. The hadronic partial decay widths of the four mixed states are computed in the $^3P_0$ model, and some branching fraction ratios are given.

Based on mass spectra and hadronic decay analyses, $D^*_J(2600)$ and $D(2750)$ are impossibly the mixed $D$ mesons of $2^3S_1$ and $1^3D_1$ at a tiny mixing angle $\theta\approx 0.12^\circ$, $D^*_{s1}(2700)$ and $D^*_{s1}(2860)$ are impossibly the mixed $D_s$ mesons of $2^3S_1$ and $1^3D_1$ at $\theta\approx 0.18^\circ$ either.

In order to identify $D^*_J(2600)$, $D(2750)$, $D^*_{s1}(2700)$ and $D^*_{s1}(2860)$, it is important to fix the accurate masses and $J^P$ numbers both from inclusive $e^+e^-$ and $pp$ collisions, and from exclusive B decays in experiment. So far, the resolve of $D^*(2760)$ and $D^*_{sJ}(2860)$ is not sufficient for the identification of $D^{*H}_1$ and $D^{*H}_{s1}$. In fact, the mass and decay data of $D^{*H}_1$ and $D^{*H}_{s1}$ has not been definitely fixed in experiments.

If the mixing angles turn large, the masses of $D^{*L}_1$ and $D^{*L}_{s1}$ turn smaller, and the masses of $D^{*H}_1$ and $D^{*H}_{s1}$ turn larger as shown in Figure. 1. Furthermore, as illustrated in Refs.~\cite{swanson,zhang1,chen}, the predicted decay widths and relevant branching ratios of the four mixed mesons would be consistent with the observed ones of $D^*_J(2600)$, $D(2750)$, $D^*_{s1}(2700)$ and $D^*_{s1}(2860)$. In this case, the problem is which kind of off-diagonal interaction can bring in a large mixing, which requires further exploration.

As pointed out in Ref.~\cite{cornell}, the leptonic or electronic decay width is more sensitive to the $^3S_1$ and $^3D_1$ mixing detail. The measurement of the leptonic or electronic decay widths will be helpful to the understanding of the dynamical mechanism of the mixing and the observed mixed states.

\begin{acknowledgments}
This work is supported by National Natural Science Foundation of China under the grants: 11975146 and 11847225.
\end{acknowledgments}

\end{document}